\documentclass[sn-mathphys]{sn-jnl_Saito}

\jyear{2023}%

\theoremstyle{thmstyleone}%
\theoremstyle{thmstyletwo}%

\theoremstyle{thmstylethree}%

\usepackage{bm} 

\begin{document}

\title[Topology of Coherent Photons]
{Representation Theory and Topology of Coherent Photons with Angular Momentum}



\author{\fnm{Shinichi} \sur{Saito}}\email{shinichi.saito.qt@hitachi.com}

\affil{\orgdiv{Center for Exploratory Research Laboratory, Research \& Development Group}, \orgname{Hitachi, Ltd.}, \orgaddress{\street{1-280 Higashikoigakubo}, \city{Kokubunji}, \postcode{185-8601}, \state{Tokyo}, \country{Japan}}}


\abstract{
Photons are elementary particles of lights \cite{Plank00,Einstein05}, which have both spin \cite{Stokes51,Jones41,Fano54} and orbital angular momentum \cite{Allen92,Padgett99,Allen00,Golub07,Bliokh09,Milione11,Bliokh15,Barnett16,Barnett16b,Bliokh17b,Moreau19} as internal degrees of freedom.
Nature of spin is known as polarisation \cite{Stokes51,Jones41,Fano54,Sotto18}, which is widely used for sunglasses, liquid-crystal displays, digital-coherent communications \cite{Bull04,Goi14,Doerr15,Kikuchi16}, while orbital angular momentum is useful for optical tweezers, laser-patterning, and quantum optics \cite{Guan13,Sotto18b,Sotto19,Abdul15b,Devlin18,Saito21f,Angelsky21,Andrews21}. 
However, spin and orbital angular momentum of photons are considered to be impossible for splitting into two independent degrees of freedom in a proper gauge invariant way, proved by plane wave expansions \cite{Enk94} in a free space \cite{Chen08,Leader14,Bliokh15,Barnett16}.
Here, we show these degrees of freedom are well-defined quantum observables in a waveguide and a free space as far as the propagation mode is sufficiently confined in the core.
We found Stokes parameters are spin expectation values of coherent photons, which exhibit non-trivial topological features \cite{Lu14,Price22} like a torus \cite{Zdagkas22}, a M\"obius strip \cite{Cisowski22}, and a bosonic Dirac cone \cite{Kumar20}.
We have applied an SU(N) representation theory to describe both spin and orbital angular momentum of photons, and experimentally demonstrated their controls over a full Poincar\'e sphere to show a fullerene C$_{60}$ \cite{Kroto85} and the earth by qubits.
We have also ascribed topological colour charge to photonic orbital angular momentum, whose SU(3) states \cite{Gell-Mann61,Gell-Mann64,Ne'eman61} are shown on a proposed Gell-Mann hypersphere in SO(8), whose parameters could be embedded in SO(5). 
We have also realised photonic SU(4) states of singlet and triplet states, which were successfully projected into SU(2)$\times$SU(2) states by a rotated polariser. 
Our results indicate that our platform of manipulating spin and orbital angular momentum is useful for exploring a photonic quantum chromodynamics and a higher order macroscopic quantum state.
}




\maketitle


One of the most intriguing motivation to develop quantum technologies is to understand why our macroscopic world is governed by classical-mechanics, even though the fundamental law of physics for its microscopic constituent is following to quantum-mechanics \cite{Caldeira81,Leggett85,Leggett95,Frowis18}.
Most of macroscopic matters cannot behave like an elementary particle, which can be a superposition state among orthogonal states, as famous for a paradox of Schr\"odinger's cat \cite{Caldeira81,Leggett85,Leggett95,Frowis18}.
However, there are a few exceptional systems, where quantum coherence is observed in a  macroscopic scale, such as superconductivity \cite{Bardeen57,Anderson58,Bogoljubov58,Nozieres85}, Bose-Einstein Condensation (BEC)\cite{Yang62,Ketterle02,Cornell02}, and a laser 
\cite{Stokes51,Jones41,Fano54}. 
In these systems, continuous symmetry is spontaneously broken upon onsets of phase transitions \cite{Bardeen57,Anderson58,Bogoljubov58,Nozieres85,Yang62,Ketterle02,Cornell02}
 or pumping above lasing thresholds, and the entire systems are described by sole wavefunctions.
For coherent photons emitted from widely available laser sources, however, it is generally believed that the polarisation state is classical, characterised by Stokes parameters \cite{Stokes51} on a topologically trivial Poincar\'e sphere \cite{Stokes51}, while these parameters are calculated quantum-mechanically using the 2 level systems with special unitary of 2-dimensions, known as SU(2) \cite{Jones41,Fano54}.

Here, we show this apparent contradiction on polarisation whether it is classical or quantum-mechanical is solved by employing a quantum many-body theory together with a representation theory of Lie algebra and Lie group.
We show coherent photons are described by a macroscopic wavefunction of SU($N$) states with spin and orbital angular momentum as internal degrees of freedom.
Based on the SU($N$) theory for coherent photons, we have experimentally confirmed that the phases and the amplitudes of the macroscopic wavefunction could be controlled by combinations of standard optical components, and the expectation values for generators of rotations were observed as Stokes and even higher order parameters.

\section{Relationship between SU($N$) states and expectation values in SO($N^2-1$)}\label{sec2}

We consider a ray of coherent photons with SU($N$) symmetry for angular momentum, which is given by the wavefunction for a coherent state \cite{Tomonaga47,Lee53,Bardeen57,Anderson58} , 
\begin{eqnarray}
\lvert \alpha_1, \cdots, \alpha_N \rangle
&=&
\prod_{\sigma=1}^{N}
{\rm e}^{-\frac{\lvert \alpha_\sigma \rvert^2}{2}}
{\rm e}^{ \hat{a}_\sigma^{\dagger} \alpha_\sigma}
\lvert 0 \rangle,
\end{eqnarray}
where $\hat{a}_\sigma^{\dagger}$ ($\hat{a}_\sigma$) is a photon creation (annihilation) operator, which satisfies the Bose commutation relationship of $[\hat{a}_\sigma, \hat{a}_{\sigma^{\prime}}^{ \dagger}]=\delta_{\sigma,\sigma^{\prime}}$,  
$\sigma$ stands for the $\sigma$-th component of the SU($N$) degrees of freedom for spin and orbital angular momentum, and $\alpha_\sigma$ is a complex number ($\mathbb{C}$) to represent the macroscopic wavefunction.
The SU($N$) degrees of freedom is related to the rotational symmetry among $N$ orthogonal states in the Hilbert space, whose rotation is given by the exponential map of the $\mathfrak{su}(N)$ Lie algebra,
\begin{eqnarray}
{\mathcal D}_i
(\theta)
&=&
{\rm e}^{-i {X}_i \theta}
,
\end{eqnarray}
where ${X}_i$ is the generator of rotation, made of a complex matrix of $N \times N$, which satisfies the commutation relationship of $[ {X}_i, {X}_j ] = i \sum_{k} f_{i j k} {X}_k$, with the structure constant of $f_{i j k}$, for the amount of the rotation, $\theta$, along the $i$-th axis.
In general, there are $(N^2 - 1)$ generators for $\mathfrak{su}(N)$, such that $i, j, k=1, \cdots, (N^2 - 1)$.
${\mathcal D}_i (\theta)$ is also a complex matrix of $N \times N$, which transfers 
$\hat{a}_\sigma^{\dagger}$ to 
$\sum_{\sigma^{\prime}} \hat{a}_{\sigma^{\prime}}^{\dagger}
{\mathcal D}_{i \sigma^{\prime} \sigma } (\theta)$, upon the rotation, or equivalently, it changes the macroscopic wavefunction in the initial state, $\lvert {\rm I} \rangle = (\alpha_1, \cdots, \alpha_N)^{\rm t}$ to the final state $\lvert {\rm F} \rangle = {\mathcal D}_i (\theta) \lvert {\rm I} \rangle$, where $^{\rm t}$ stands for the transpose.

For spin angular momentum of coherent photons, we use SU(2) states for horizontal-vertical (HV) linearly-polarised or left-right (LR) circularly-polarised states as fundamental bases \cite{Stokes51,Jones41,Fano54}, and the generators of rotations are given by Pauli matrices, ${\sigma}_i$, $i=1,2,3$, which satisfy $[ {\sigma}_i, {\sigma}_j ] = 2 i \sum_{k} \epsilon_{i j k} {\sigma}_k$, where $\epsilon_{i j k} $ is the totally asymmetric tensor.
We have obtained the spin angular momentum operator in the many-body state, as $\hat{\bf S}=\hbar \hat{\bm \psi}^{\dagger} {\bm \sigma} \hat{\bm \psi}$, where $\hat{\bm \psi}^{\dagger}=(\hat{a}_{\rm H}^{\dagger} , \hat{a}_{\rm V}^{\dagger} )$ and $\hat{\bm \psi}=(\hat{a}_{\rm H} , \hat{a}_{\rm V} )^{\rm t}$ are the spinor representation of creation and annihilation operators in HV bases to create and annihilate a photon, and ${\bm \sigma} = (\sigma_3, \sigma_1, \sigma_2)$ are generators of spin angular momentum.
In the HV bases, the macroscopic wavefunction is given by $(\alpha_{\rm H},\alpha_{\rm V})=\sqrt{\mathcal N}(\cos \alpha, \sin \alpha {\rm e}^{i \delta})$, where $ {\mathcal N}$ is the average density of photons passing the cross section of the waveguide per second, $\alpha$ is the auxiliary angle, and $\delta$ is the phase.
Then, it is straightforward to calculate the quantum-mechanical average of $\hat{\bf S}$, since the coherent state is the eigenstate of $\hat{\bm \psi}$, and we obtain $\langle \hat{\bf S} \rangle_{\rm I}= \hbar {\mathcal N}(\cos(2\alpha), \sin(2\alpha) \cos \delta , \sin(2\alpha) \sin \delta)$, where $\hbar$ is the Dirac constant.
$\langle \hat{\bf S} \rangle_{\rm I}$ is equivalent to normalised Stokes parameters of ${\bf S}=(S_1, S_2, S_3)$, but we also obtained the overall magnitude, $S_0=\hbar {\mathcal N}$, which means the Plank constant ($h=2 \pi \hbar$) effectively becomes macroscopic for coherent photons.

Moreover, above relationship between SU($N$) states and generators of rotation leads to a general formula for expectation values in the final state upon the rotation of $\theta$ as 
\begin{eqnarray}
\langle 
{X}_j
\rangle_{\rm F}
&=&
\lim_{n\rightarrow \infty}
\sum_k
\left( 1- {F}_{i} \frac{\theta}{n} \right)_{jk}^{n}
\langle {X}_k \rangle_{\rm I}
=
\sum_k
\left( 
{\rm e}^{- {F}_{i} \theta}
\right)_{jk}
\langle {X}_k \rangle_{\rm I}
,
\end{eqnarray}
where ${F}_{i}$ is an adjoint operator, whose matrix element becomes $({F}_{i})_{jk}=f_{ijk}$, which is a matrix of $(N^2-1)\times(N^2-1)$.
This means that the rotation of the wavefunction in the Hilbert space of SU($N$) corresponds to the rotation of expectation values in the special orthogonal group of SO($N^2-1$).
For spin angular momentum of SU(2), this corresponds to the rotation of Stokes parameters on the Poincar\'e sphere in SO(3).

Now, we understand why Stokes parameters on Poincar\'e sphere can describe coherent photons, emitted from a laser.
The original rotational symmetry of the system is broken upon the lasing threshold, and the single mode (or a few modes, depending on the quality of the cavity) is naturally selected upon stimulated emissions.
Thus, the whole system is described by a sole macroscopic wavefunction, which has internal degrees of freedom for spin angular momentum, and the existence of the Nambu-Anderson-Higgs-Goldstone mode upon symmetry-breaking guarantees that we can still rotate the whole wavefunction without spending energies 
\cite{Higgs64,Goldstone62,Nambu59}.
Consequently, we can manipulate the polarisation state by using wave-plates and phase-shifters, which work as quantum-mechanical rotation operators in SU(2), ${\mathcal D}_i(\theta)={\rm e}^{-i {\sigma}_i  \theta/2}$.
We also understand the inherent relationship between angular momentum states \cite{Stokes51,Jones41,Fano54}  and a representation theory \cite{Cisowski22}.

\section{Poincar\'e rotator}\label{sec2}
Our next step is to achieve an arbitrary SU(2) rotations for coherent photons in experiments.
It is generally well-known that the rotated half-wave-plate (HWP) becomes a pseudo-rotator rather than a pristine rotator.
Thus, it was required to know the SU(2) state of the input in order to realise the intended rotation of ${\bf S}$, but the measurement projects the SU(2) state and a quantum operation should be applicable without knowing the state.
In fact, the rotated HWP corresponds to the Mirror reflection, which belongs to a set of O$^{-}(2)$, whose determinant is -1, and O$^{-}(2)$ does not form a proper group as evident from the fact that twice mirror reflections result in an identity operation, ${\bf 1}$, rather than the product of rotations.
This could be overcome just to prepare another HWP, whose fast axis (FA) is fixed to a certain direction, before applying the rotated HWP.
The operations of 2 HWPs correspond to a proper rotation to form SO(2)=O$^{+}(2)$, whose determinant is 1.
For the wavefunction, this corresponds to apply $U(1)$ operation, which is evident for the LR bases, since the corresponding wavefunction of $(\alpha_{\rm L},\alpha_{\rm R})=({\rm e}^{-i \phi /2} \cos(\theta/2),{\rm e}^{i \phi /2} \sin(\theta/2))$ is rotated to change the azimuthal angle $\phi$ upon the rotation, while the polar angle of $\theta$ is preserved.
Thus, we can realise an arbitrary amount of rotations along the $S_3$ axis on the Poincar\'e sphere, simply by rotating HWP together with a fixed HWP.
We can also realise an arbitrary amount of the phase-shift, just by adding 2 quarter-wave-plates (QWPs) before and after the rotator.
We should align the FA of the first QWP to the anti-diagonal direction to convert the $S_2$-$S_3$ plane to the $S_1$-$S_2$ plane to apply a rotator operation by 2 HWPs, and then aligh the FA of the last QWP to the diagonal direction for bringing the axis back to the original axis.
Consequently, we could realise a phase-shifter to allow an arbitrary rotation of ${\bf S}$ along the $S_1$ axis.
We propose to call this device as a Poincar\'e rotator, since we can realise an arbitrary SU(2) operation for the polarisation state.
We confirmed the expected passive operations, and we could also realise an active Poincar\'e rotator by using optical modulators for the $U(1)$ phase-shifts rather than physical rotations.

\begin{figure}[h]%
\centering
\includegraphics[width=0.9\textwidth]{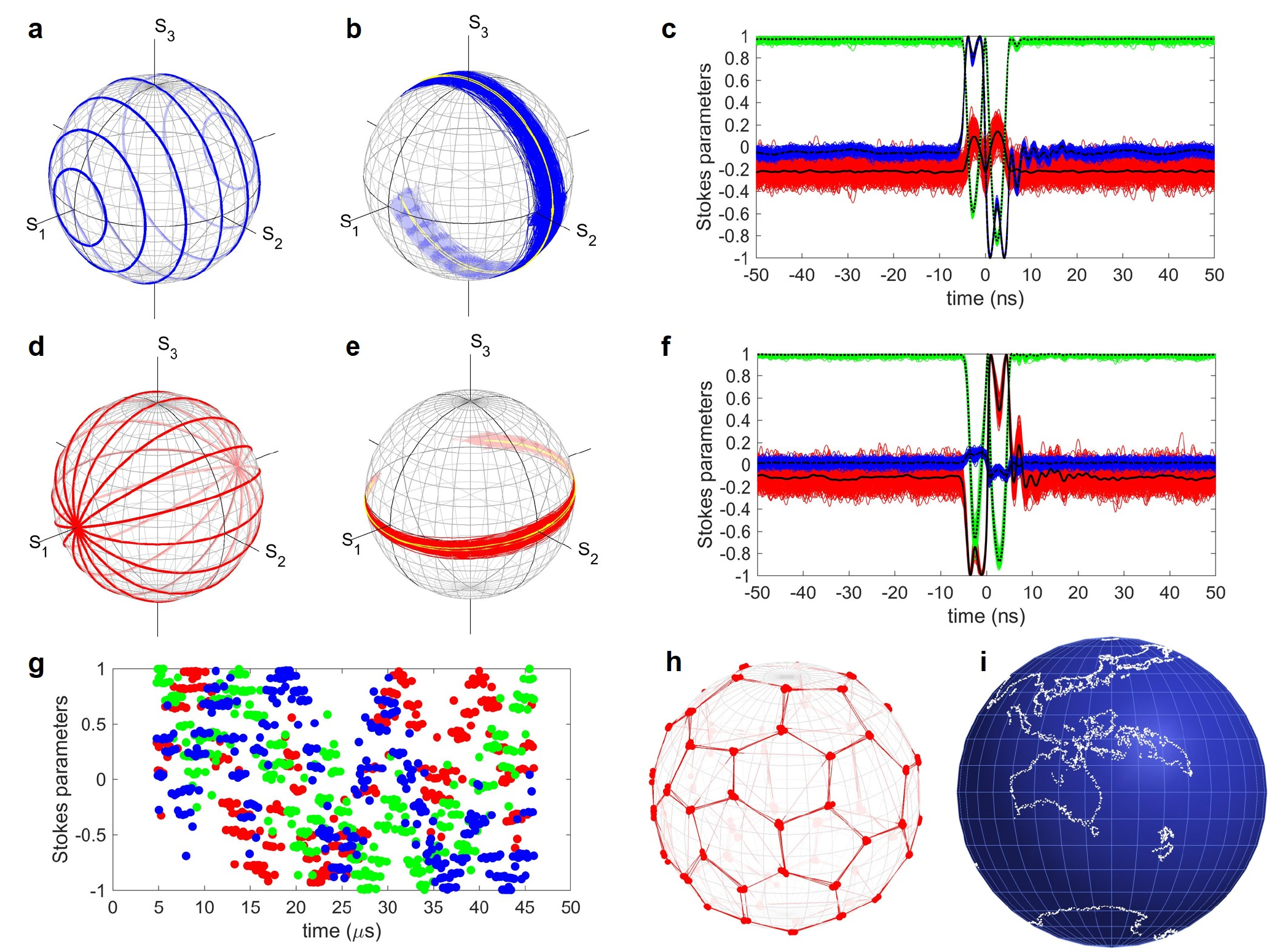}
\caption{
Poincar\'e rotator to realise an arbitrary SU(2) operation for spin states of coherent photons, using a laser with the wavelength of 1533nm.
(a) and (d) simulated trajectories for phase-shifter and rotator operations, respectively.
(b) and (c) phase-shifter operations, and (e) and (f) rotator operations, obtained by the input voltage of 4.6V applied to optical modulators. 
(g) and (h) fullerene C$_{60}$, and (i) the earth corresponding to 9,865 points, operated at 5 MHz with the duty cycle of 5 \%.
Stokes parameters of $S_1$ (red), $S_2$ (green), and $S_3$ (blue) are shown in (c), (f), and (g).
}\label{fig1}
\end{figure}

The active Poincar\'e rotator operations are shown in Fig. 1.
The phase-shifter operations were confirmed (Fig. 1 (a)-(c)) simply by splitting HV states using a polarisation-beam-splitter (PBS), followed by the phase-shifts of LiNbO$_3$ optical modulators, and recombined by a polarisation-beam-combiner (PBC).
The amplitudes of HV states were controlled by the rotator operations (Fig. 1 (d)-(f)) by using another phase-shifter together with 2 QWPs.
At this moment, the fidelity is as low as $\sim 90$ \%, which was limited by expected deviations of rotation angles in achromatic wave-plates. 
By combining the phase-shifter and the rotator, we could realise arbitrary SU(2) states of polarisation, which corresponds to a single qubit operation.
It is worth for noting that the rotation axis is rotated for the rotator operation of Fig. 1 (d) upon the phase-shifts, while the phase-shifter operation is always along the $S_1$ axis due to the splitting by the PBS at the fixed directions.
This naturally allows us to control both polar and azimuthal angles on the Poincar\'e sphere.
As a demonstration of the Poincar\'e rotator, we have tried to draw the fullerene C$_{60}$ (Fig. 1 (g) and (h)) \cite{Kroto85} and the coastline of the earth (Fig. 1 (i)) on the Poincre\'e sphere.
The operation speed of 5 MHz was limited by the performance of electronics, used for this study, such as arbitrary-wave-form generators and amplifiers.

\section{Topological polarisation states}

\begin{figure}[h]%
\centering
\includegraphics[width=0.9\textwidth]{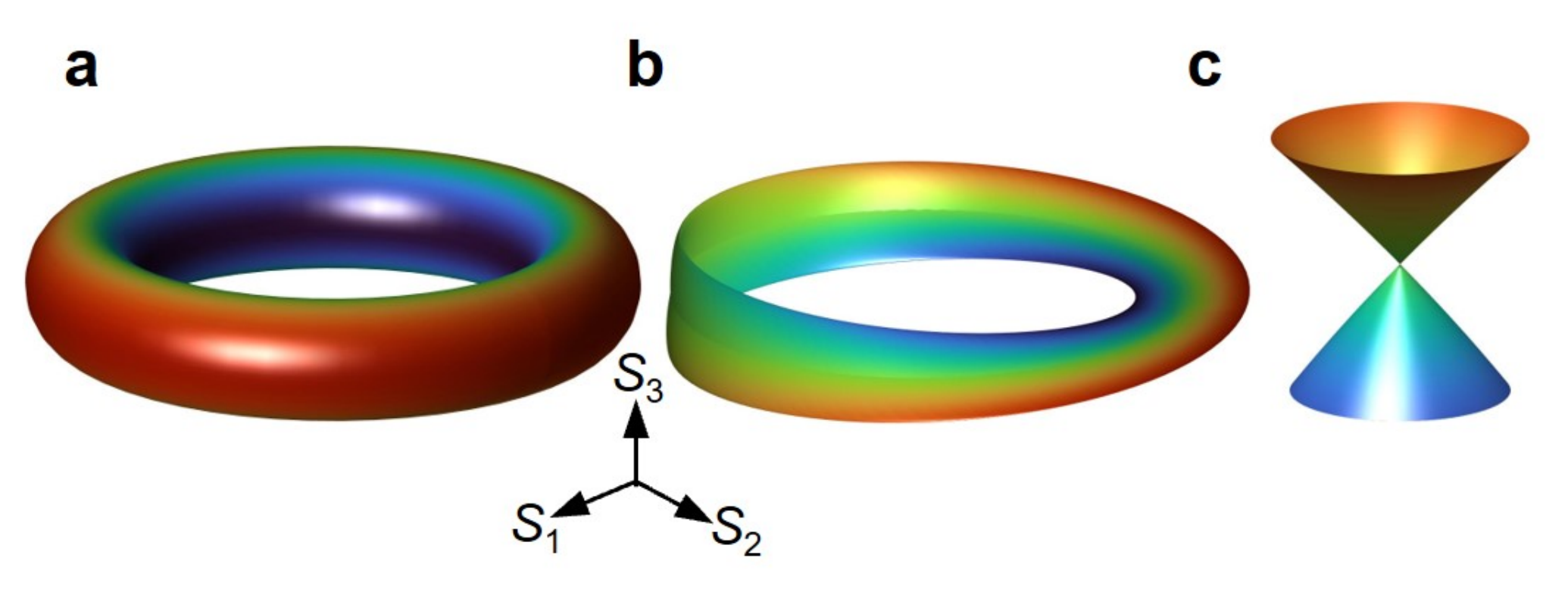}
\caption{
Topological polarisation states.
Stokes parameters were theoretically calculated for (a) a torus, (b) a M\"obius strip, and (c) a bosonic Dirac cone.
}\label{fig2}
\end{figure}

Regardless of the demonstrations for SU(2) operations to coherent photons from a laser source \cite{Stokes51,Jones41,Fano54}, one might still think that the state is considered to be classical, since it is characterised by a real observable of ${\bf S}$ on a sphere of ${\mathbb S}^2$, which is topologically trivial in SO(3).
There is a difference, however, between SU(2) and SO(3), since SU(2) is a two-fold coverage group for SO(3), shown as ${\rm SU(2)}/{\mathbb S}^{0} \cong {\rm SO(3)}$, where ${\mathbb S}^{0}= \{ -1, 1 \} $ is the zero-dimensional sphere, corresponding to the parity of rotations.
Specifically, this yields ${\mathcal D}_i(2\pi)=-{\bf 1}$ for coherent photons, whose phase-change for the output after one rotational operation can be observed upon the interference to the original input state, independent on the rotational axis of $i$.
We have also realised ${\bf S}$ can fully span the three-dimensional Euclidean space, which we call Stokes space, since we are dealing with the macroscopic number of photons due to Bose-Einstein statistics, and the radius of the Poincar\'e sphere corresponds to ${\mathcal N}$.
Consequently, we can realise topologically non-trivial structures \cite{Lu14,Price22} as trajectories of ${\bf S}$.

Examples of topological polarisation states are shown in Fig. 2. 
The torus \cite{Zdagkas22} of ${\mathbb T}^2 \cong {\mathbb S}^1 \times {\mathbb S}^1$ can be realised by a polarisation independent splitter and a combiner, together with a phase-shifter and a rotator.
The splitter separates the input ray into 2 rays, and one of the ray is phase-shifted by HWPs, while the other ray is kept its polarisation state, and then they are recombined at the combiner.
This simple set-up allows us to realise the interference to change the amplitude for ${\mathcal N}$, which changes the radius of the trajectory to form a circle of $ {\mathbb S}^1$.
Then, we can subsequently rotate the polarisation state along the direction perpendicular to the radius, to form a torus in the Stokes space (Fig. 2 (a)).
We have experimentally confirmed the torus, in a fibre-optic polarisation interferometer, together with passive optical components to use HWPs and QWPs.
We can also realise more complex structures, such as a M\"obius strip (Fig. 2 (b)) \cite{Cisowski22} and a bosonic Dirac cone (Fig. 2 (c)) \cite{Kumar20}, whose node must exist at the edge to change from a topologically non-trivial torus to a topologically trivial sphere, satisfying the Gauss-Bonnet theorem \cite{Chern46} to account for the bulk-edge correspondence. 

\section{SU(3) states and Gell-Mann hypersphere}

We have theoretically confirmed that we can split spin and orbital angular momentum, as far as the ray is predominantly propagating along one direction ($z$).
In this case, the ray must have a mode profile in the $xy$ plane, which could be Gaussian, Laguerre-Gauss, or whatever, and consequently the mode has a small longitudinal component.
By considering the longitudinal component, we could derive the helical components of spin and orbital angular momentum in a proper gauge invariant manner.
Assuming the rotational symmetry of angular momentum, we could obtain both spin and orbital angular momentum operators exactly for a graded-index (GRIN) waveguide \cite{Kawakami68} and approximately for a free space under the paraxial approximation.
We also confirmed the ladder operator works properly for orbital angular momentum, similar to the 
Cartan-Dynkin scheme for Lie algebra, which can be experimentally realised by a vortex lens \cite{Golub07,Abdul15b}.

\begin{figure}[h]%
\centering
\includegraphics[width=0.9\textwidth]{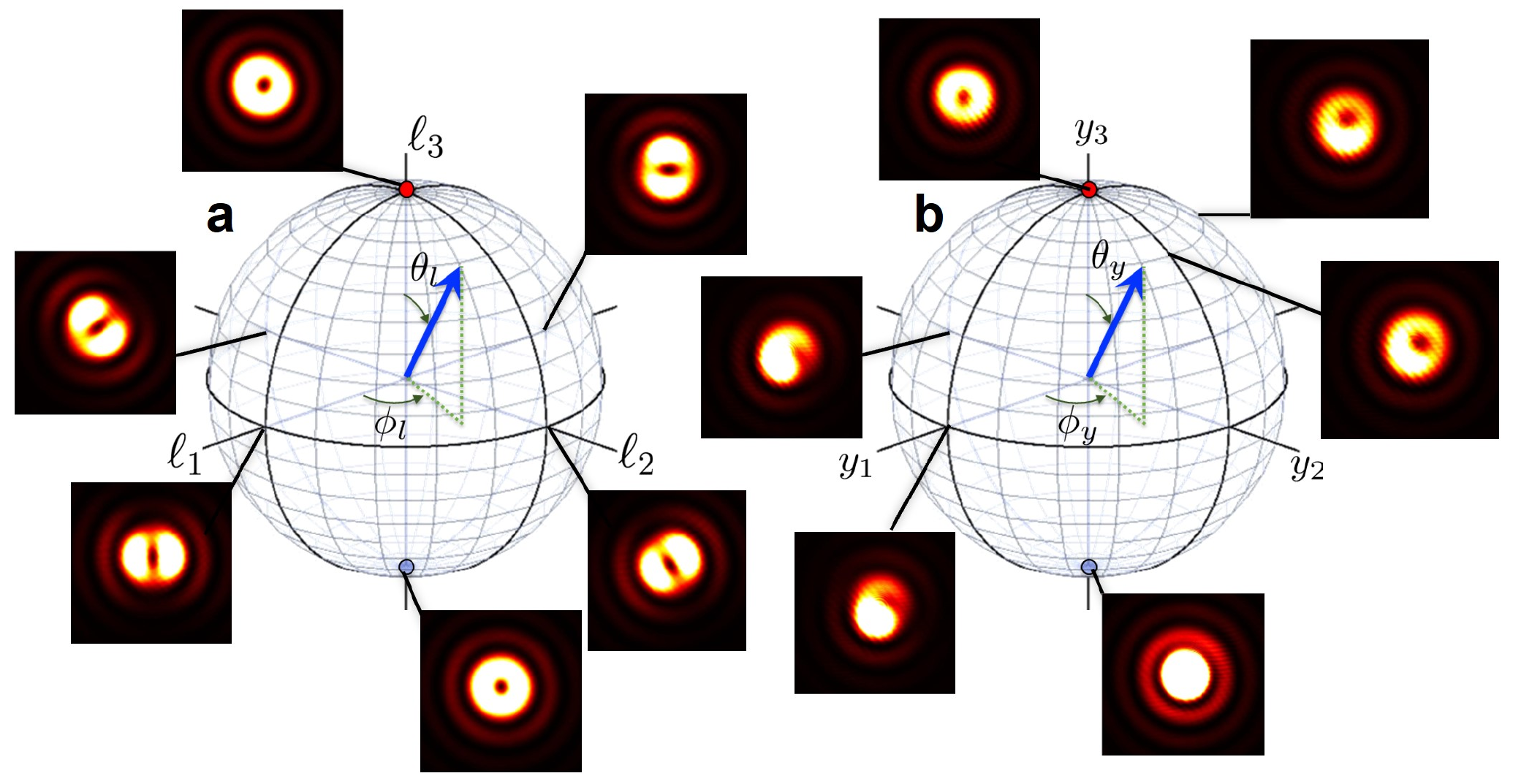}
\caption{
Gell-Mann parameters to characterise the SU(3) states of coherent photons, which are realised by superposition states of left- and right-vortexed states and a Gaussian state.
(a) Orbital angular momentum, ${\bm \ell}=(\ell_1,\ell_2,\ell_3)$, and (b) hyperspin, ${\bf y}=(y_1 , y_2 , y_3)$.
A laser with the wavelength of 532nm was used and far-field images were taken.
}\label{fig3}
\end{figure}

Our next step to explore a macroscopic quantum nature of coherent photons is to realise the higher order SU(3) state by using orbital angular momentum \cite{Allen92,Padgett99,Allen00,Golub07,Bliokh09,Milione11,Bliokh15,Barnett16,Barnett16b,Bliokh17b,Moreau19}.
We consider left- and right-vortexed states, which has topological charge of $1$ and $-1$ at the centre of the modes, respectively, together with a standard Gaussian mode without a vortex.
We assign SU(3) states for those orthogonal states, which are clearly distinguishable by topological colour charge, similar to quarks \cite{Gell-Mann61,Gell-Mann64,Ne'eman61}, while we consider the polarisation state is the same.

The SU(3) states are characterised by 8 generators of rotations \cite{Gell-Mann61,Gell-Mann64,Ne'eman61}, whose quantum-mechanical expectation values should be observable, since generators of rotations are Hermite.
We propose to call these parameters, as Gell-Mann parameters, named after Murray Gell-Mann, who discovered quarks.
As shown in Eq. (3), Gell-Mann parameters can be shown on a hypersphere of 8-dimensions in SO(8), whose radius is normalised to be $2/\sqrt{3}$, since the Casimir operator is constant.
We have analysed Gell-Mann parameters, and found that we do not need all 8 parameters to describe SU(3) states, which could be described on ${\mathbb C}^3$ under the normalisation, while the global phase is not required, such that 4 real parameters on ${\mathbb S}^4$ is enough, embedding in SO(5).
Practically, it would be useful to show Gell-Mann parameters, using 2 Poincar\'e spheres (Fig. 3); one for orbital angular momentum (Fig. 3 (a)) \cite{Padgett99,Milione11} and the other for hyperspin (Fig. 3 (b)), which we defined to describe coupling between vortexed modes and a Gaussian mode.

We have experimentally realised such superposition states of SU(3), and obtained the far-field images (Fig. 3).
We have employed the Poincar\'e rotator for polarisation to prepare rays of expected amplitudes and phases, and subsequently used vortex lenses to generate both left- and right-vortices and recombined to recover the quantum coherence as 1 ray.
This scheme worked properly as a spin-to-orbit converter, and we have realised expected superposition states of left- and right-vortices (Fig. 3 (a)) \cite{Padgett99,Milione11}.
Moreover, we have realised a superposition state with and without the vortices, which would be fundamentally of interest, since a doughnut cannot be a ball by topological deformation (Fig. 3 (b)).
We found that the superposition state is characterised by the motion of the topological charge, which could leave from the centre of the mode upon changing the amplitude, and the direction of the motion could be controlled by the phase-shift, using our Poincar\'e rotator.

\section{Macroscopic singlet and triplet}

\begin{figure}[h]%
\centering
\includegraphics[width=0.9\textwidth]{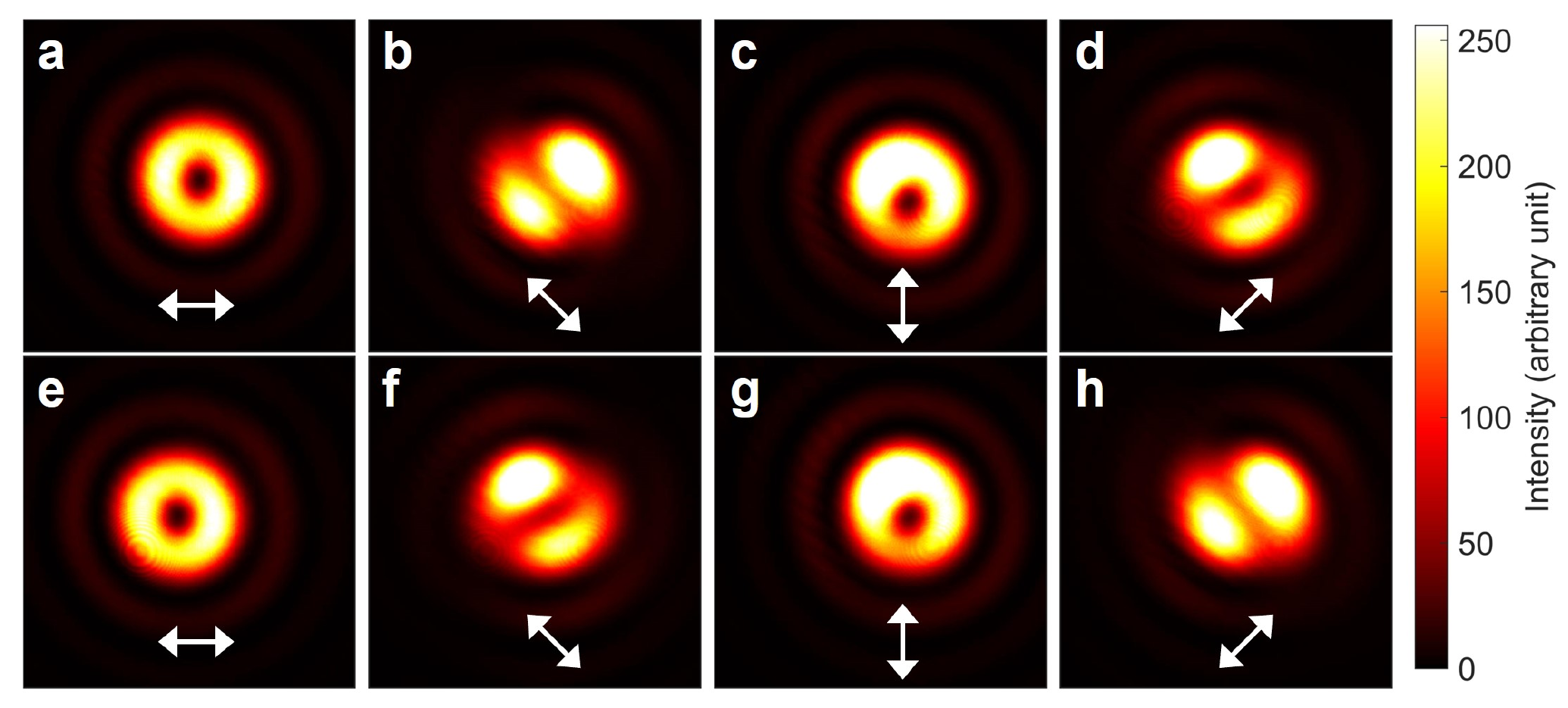}
\caption{
Macroscopic singlet and triplet states of coherent photons, from a laser with the wavelength of 532nm.
Far-field images were taken for (a)-(d) singlet states, and (e)-(h) triplet states.
Arrows indicate the projected direction for the polarisation.
The chiralities are not obvious for (a) and (e) right vortexed state, and (c) and (g) left vortexed state.
On the other hand, the directions of dipoles are clearly perpendicular to the polarisation for singlet states of (b) and (d), while the directions of dipoles are aligned to the polarisation for triplet states of (f) and (h).
}\label{fig4}
\end{figure}

The final challenge for this paper is to realise macroscopic singlet and triplet states by using both spin and orbital angular momentum of coherent photons.
As we have confirmed both theoretically and experimentally, coherent photons have both spin and orbital angular momentum, which at least forms SU(2)$\times$SU(2) states, as far as they are independently controlled.
If we distinguish all 4 orthogonal states, 
$\lvert \uparrow     \rangle_{\rm s} \lvert \uparrow     \rangle_{\rm o}$, 
$\lvert \uparrow     \rangle_{\rm s} \lvert \downarrow \rangle_{\rm o}$, 
$\lvert \downarrow \rangle_{\rm s} \lvert \uparrow     \rangle_{\rm o}$, 
and 
$\lvert \downarrow \rangle_{\rm s} \lvert \downarrow \rangle_{\rm o}$,
for both spin and orbital angular momentum, 
we should be able to realise SU(4) states.
Here, we have assigned spin up and down states to horizontal and vertical states for polarisation, respectively, while up and down states of the orbital angular momentum correspond to left and right vortexed states, respectively.

The singlet state is defined by $\lvert {\rm Singlet} \rangle = (\lvert \uparrow     \rangle_{\rm s} \lvert \downarrow \rangle_{\rm o} - \lvert \downarrow \rangle_{\rm s} \lvert \uparrow     \rangle_{\rm o})/\sqrt{2}$, while the triplet state is $\lvert {\rm Triplet} \rangle = (\lvert \uparrow     \rangle_{\rm s} \lvert \downarrow \rangle_{\rm o} + \lvert \downarrow \rangle_{\rm s} \lvert \uparrow     \rangle_{\rm o})/\sqrt{2}$.
We have experimentally realised these states by using Poincar\'e rotator together with the vortex lenses.
In order to confirm the realisations of these states, we have projected the SU(4) states into SU(2)$\times$SU(2) states by a linear polariser (Fig. 4).
Upon passing through the polariser, the polarisation state aligned to the direction of the polarisation was selected, and consequently, the state for orbital angular momentum was also selected. 
For example, if $\lvert {\rm Singlet} \rangle$ is projected by the horizontally polarised state, the projected state became $\lvert \uparrow     \rangle_{\rm s} \lvert \downarrow \rangle_{\rm o}/\sqrt{2}$, which must be in the right vortexed state (Fig. 5 (a)).
Unfortunately, we cannot distinguish the left-vortexed state (Fig. 5 (c)) with the right-vortexed state (Fig. 5 (a)) from the images, for example, since the chirality of the vortex disappeared in intensities. 
This issue could be overcome, if we have projected the polarisation state by the diagonal (Fig. 5 (d) and (h)) and anti-diagonal polarisers (Fig. 5 (b) and (f)).
In this case, dipole images of orbital angular momentum states must be aligned along the direction of orbital angular momentum.
If the state is singlet, the total angular momentum must vanish, such that directions of the dipoles must be perpendicular to those of spin, aligned to the polariser  (Fig. 5 (b) and (d)).
On the contrary, if the state is triplet, the dipoles should be aligned to the same directions with spin (Fig. 5 (f) and (h)).
We could successfully observe expected behaviours, which are evidences for the realisation of macroscopic singlet and triplet states.


\section{Conclusion}\label{sec13}

In conclusion, we have theoretically shown that a macroscopic coherent state with an SU($N$) symmetry gives expectation values of its generators of rotations, shown in SO($N^2-1$).
Based on the representation theory, we have experimentally realised a Poincar\'e rotator, topological polarisation states, colour-charged states, and macroscopic singlet and triplet states.
Our hypothesis was coherent states of photons, emitted from a conventional laser, are described by a macroscopic wavefunction, to exhibit quantum-mechanical behaviours in a macroscopic scale.
Our results are so far consistent with the view, that coherent photons have at least a few internal degrees of freedom as angular momentum, and a standard quantum-mechanical framework is applicable to understand its expectation values and coupling between spin and orbital angular momentum.
It will be useful to use our experimental platform to explore a higher order quantum state such as SU(6) and its projection to SU(2)$\times$SU(3) for photonic quantum chromodynamics and beyond.

\bmhead{Acknowledgments}

This work is supported by JSPS KAKENHI Grant Number JP 18K19958.
The author would like to express sincere thanks to Prof I. Tomita for continuous discussions and encouragements.

\bibliography{Saito}

\end{document}